\documentclass[a4paper,11pt]{article}
\usepackage{pos}
\usepackage{slashed}

\title{Phenomenology of Spin-1 resonances at the LHC}

\author*{Werner Porod}

\affiliation{University of W\"urzburg, Institute of Theoretical Physics and Astrophysics \\
        Campus Hubland Nord, Emil-Hilb-Weg 22, D-97074 W\"urzburg, Germany}

\emailAdd{werner.porod@uni-wuerzburg.de}

\abstract{Spin-1 resonances are among the states predicted by composite Higgs models and one
expects them the have masses in the range of a few TeV. We focus here on 
models based on an underlying gauge-fermion description which predict QCD-coloured vector and axial-vector states as well as states charged under the electroweak gauge groups. The former can come as  triplet, sextet and octet representation depending on the model details. All models
considered have a colour octet vector state in common which can be singly
produced at hadron colliders as it mixes with the gluon.  We summarize
here their LHC phenomenology and comment also on aspects relevant for future colliders.
}

\FullConference{Proceedings of the Corfu Summer Institute 2024 "School and Workshops on Elementary Particle Physics and Gravity" (CORFU2024)\
12 - 26 May, and 25 August - 27 September, 2024\\
Corfu, Greece\\}

\newcommand{\ii}{\mathrm{i}}
\newcommand{\Tr}{\mathrm{tr}}

\begin{document}
\maketitle

\section{Introduction}

Composite Higgs models  \cite{Kaplan:1983fs,Kaplan:1983sm} postulate the existence of an additional gauge interaction which
becomes strong in the multi-TeV range. In this way, these models provide a potential 
solution to the problem of hierarchy between the electroweak scale and the Planck scale: 
like in quantum chromodynamics (QCD), the breaking scale is dynamically generated via 
confinement and condensation of the new interaction. The Higgs boson emerges as 
a pseudo Nambu Goldstone boson (pNGBs) \cite{Contino:2003ve} from the spontaneous breaking 
of the global symmetry in the new strong sector. Its potential is generated by explicit
breaking terms: the gauging of the electroweak symmetry, the couplings of the top quark
\cite{Agashe:2004rs} and (eventually) a mass term for the underlying fermions \cite{Galloway:2010bp,Cacciapaglia:2014uja}.

The largeness of the top-mass can be explained via partial
compositeness  \cite{Kaplan:1991dc} which leads to the prediction of top-partners. 
The fact, that these need to carry QCD charges implies that there are also 
coloured spin-0 and spin-1 states, as well as fermions carrying unusual colour charges.
In this contribution, we focus on models based on an underlying 
gauge-fermion description as they allow for a systematic classification of the properties
of the resonances. A systematic list of models describing the minimal resonances required 
by the Higgs sector has been presented in ref.~\cite{Ferretti:2013kya}. An important
class are model models with two separate species in different irreducible representations
(irreps). A set of twelve minimal models, dubbed M1-M12, has been defined in 
ref.~\cite{Ferretti:2016upr,Belyaev:2016ftv} 
which are fully characterized in terms of the confining gauge group and the irreps and 
multiplicities of the two fermion species.

Several studies have been presented in the literature covering the phenomenology of various
resonances of these twelve models: 
electroweak pNGBs \cite{Ferretti:2016upr,Agugliaro:2018vsu,Cacciapaglia:2022bax,Flacke:2023eil},
singlets stemming from the global U(1)'s \cite{Ferretti:2016upr,Belyaev:2016ftv,Cacciapaglia:2019bqz,BuarqueFranzosi:2021kky},
QCD coloured pNGBs \cite{Cacciapaglia:2015eqa,Belyaev:2016ftv,Cacciapaglia:2020vyf}, top 
partners with non-standard decays \cite{Bizot:2018tds,Xie:2019gya,Cacciapaglia:2019zmj,Banerjee:2022izw,Banerjee:2024zvg} or 
colour assignment \cite{Cacciapaglia:2021uqh}, and spin-1 resonances carrying electroweak 
charges \cite{BuarqueFranzosi:2016ooy,Caliri:2024jdk} or QCD charges \cite{Cacciapaglia:2024wdn}.
We note for completeness, that spectra and couplings of such resonances have been
computed on the Lattice for models based on Sp(4) 
\cite{Bennett:2017kga,Bennett:2019cxd,Bennett:2019jzz,Bennett:2022yfa,Kulkarni:2022bvh,Bennett:2023mhh,Bennett:2024tex}, 
e.g.~models M5 and M8, and models based on SU(4) 
\cite{Ayyar:2017qdf,Ayyar:2018glg,Ayyar:2018ppa,Ayyar:2018zuk,Ayyar:2019exp,Hasenfratz:2023sqa}, 
e.g.~models M6 and M11. Computations based on holography are also available 
\cite{Erdmenger:2020lvq,Erdmenger:2020flu,Elander:2020nyd,Elander:2021bmt,Erdmenger:2023hkl,Erdmenger:2024dxf}. 
The subset of models covered by both approaches give consistent results.

In this contribution we focus on the phenomenology of color charged spin-1 resonances  
which are present in all the models proposed in ref.~\cite{Ferretti:2016upr,Belyaev:2016ftv}. We summarize
first relevant model aspects in the next section. We then discuss in the subsequent section to which
extent this sector is constrained by existing LHC data. The final section concludes with a discussion
and an outlook.

\section{Model aspects}

The models presented ef.~\cite{Ferretti:2016upr,Belyaev:2016ftv} propose that the hyperfermions charged
under the new gauge come in two different representations such that
 top partners emerge as so called ``chimera'' baryons. These can be realized in two different patterns: 
 $\psi \psi \chi$ and $\psi \chi \chi$, where $\psi$ only carry electroweak charges while $\chi$ carry QCD colour and hypercharge. In the former case, the $\chi$'s QCD triplet carries hypercharge $2/3$, in the
latter case $-1/3$. The spin-1 resonances carrying QCD charges  emerge as bound states of the $\chi$
species. Their properties emerge from three types of cosets, SU(6)/SO(6), SU(6)/Sp(6) and 
SU(3)$\times$SU(3)/SU(3).

The resulting spectrum contains a set of vectors $\mathcal V^\mu$ as well as a set of axial-vectors $\mathcal A^\mu$, which decay respectively into two or three pNGBs\footnote{This is actually an abuse of language. Only in case of the coset SU(3)$\times$SU(3)/SU(3) the names 'vector' and 'axial-vector´ 
coincides with the CP properties of the
spin-1 resonances.}. The latter property originates from 
the symmetric nature of the cosets \cite{Cacciapaglia:2024wdn}. 
All cosets contain an ubiquitous octet $\mathcal V^\mu_8$ within $\mathcal V^\mu$ which mixes with the 
QCD gluons. In addition there is always a color singlet within
$\mathcal V^\mu$, a color octet within $\mathcal A^\mu$ as well as a color octet pNGB $\pi_8$. The coset
SU(6)/SO(6) contains in addition a color triplet $\mathcal V^\mu_3$, a color sextet  $\mathcal A^\mu_6$
and a color sextet pNGB  $\pi_6$ whereas the SU(6)/Sp(6) contains in addition a color sextet 
$\mathcal V^\mu_6$, a color triplet  $\mathcal A^\mu_3$ and a color triplet pNGB  $\pi_3$. The electric
charges of the triplet and sextet states depend on the model details and are summarized in 
table~\ref{tab:spectrumQCD}.

\begin{table}[t]
\centering\small
\begin{tabular}{|c|c|c||c|c|c|c|c|}
\hline
& Models & $\chi$ ($R,Y,B$) & $\pi$ & $\mathcal{V}^\mu$ & $\mathcal{A}^\mu$ & $\Psi$ &di-quark\\ \hline\hline
C1 & M1-2 & (R,$-\frac{1}{3}$,$\frac{1}{6}$) & $8_0,\, {\color{red} 6_{-2/3}}$ & $8_0,\, 1_0,\, {\color{red} 3_{2/3}}$ & $8_0,\, {\color{red} 6_{-2/3}}$ & $8,\, 1,\, {\color{red} 3},\, {\color{red} 6}$ & none\\ \hline
C2 & M3-4, M8-11 & (R,$\frac{2}{3}$,$\frac{1}{3}$) & $8_0,\, {\color{blue} 6_{4/3}}$ & $8_0,\, 1_0,\, {\color{blue} 3_{-4/3}}$ & $8_0,\, {\color{blue} 6_{4/3}}$ & ${\color{red} 3}$ & $\pi_6,\mathcal{V}^\mu_3,\mathcal{A}^\mu_6$\\ \hline
C3 & M5 & (Pr,$-\frac{1}{3}$,$\frac{1}{6}$) & $8_0,\, {\color{red} 3_{2/3}}$ & $8_0,\,  1_0,\, {\color{red} 6_{-2/3}}$ & $8_0,\, {\color{red} 3_{2/3}}$ & $8,\, 1,\, {\color{red} 3},\, {\color{red} 6}$ & none\\ \hline
C4 & M6-7 & (C,$-\frac{1}{3}$,$\frac{1}{6}$) & $8_0$ & $8_0,\,  1_0$ & $8_0$ & $8,\, 1,\, {\color{red} 3},\, {\color{red} 6}$ & none\\ \hline
C5 & M12 & (C,$\frac{2}{3}$,$\frac{1}{3}$) & $8_0$ & $8_0,\,  1_0$ & $8_0$ & ${\color{red} 3}$ & none\\ \hline
\end{tabular}
\caption{\label{tab:spectrumQCD} Properties of the spin-0 ($\pi$), spin-1 ($\mathcal{V}^\mu$, 
$\mathcal{A}^\mu$) and spin-1/2 ($\Psi$) lightest resonances in the 12 models, grouped in 5 classes. 
Each class is determined by the properties of the $\chi$ species, listed in the second column by irrep type:
R for real, Pr for pseudo-real and C for complex. For the resonances, the colours indicate the baryon
numbers: black for $B=0$, red for $B=\pm 1/3$ and blue for $B=\pm 2/3$. In the last column we indicate
the bosons that can decay into a $tt$ di-quark state.}
\end{table}

In ref.~\cite{Cacciapaglia:2024wdn,Kunkel:2025qld}, to which we refer for details, the hidden symmetry 
approach \cite{Bando:1987br} has been used to obtain the effective Lagrangian for these states. In the 
following we denote the mass of the octet vector by $M_{\mathcal V_8}$, the mixing angle between
these states and the gluons by $\beta_8$ and gauge coupling of the new strong sector by $\tilde g$.
The main decay modes of the heavy spin-1 resonances are generated by three types of interactions:
\begin{itemize}
 \item couplings to pNGBs from the chiral Lagrangian in the strong sector;
 \item couplings to quarks via the mixing of the colour octet to gluons;
 \item partial compositeness couplings to top and bottom quarks.
\end{itemize}
The first type stems directly from the pNGB embedding in the effective Lagrangian. It turns out
that  only two operators contribute \cite{Cacciapaglia:2024wdn,Kunkel:2025qld}
\begin{align}
	\mathcal O_{\mathcal V} &= \ii \,\Tr([\boldsymbol \pi, \partial_\mu \boldsymbol \pi] \boldsymbol{ V}^\mu),\\
	\mathcal O_{\mathcal A} &= \Tr( [\boldsymbol \pi, [\boldsymbol \pi, \partial_\mu \boldsymbol \pi]] \boldsymbol {\mathcal A}^\mu ) \,,
\end{align}
if one requires to have at most three pNBGs at a vertex. Both, $\mathcal O_V$ and $\mathcal O_{\mathcal A}$ are hermitian. The corresponding coupling of the octet vector is given by 
\begin{align}
	C_{\mathcal V_8} &= \frac{g_{\rho\pi\pi}}{\cos\beta_8} + 2\frac{1+R^2}{1-R^2} g_s \tan\beta_8\,, \label{eq:cv8} 
\end{align}
where $g_{\rho\pi\pi}$ is the coupling of the pNGBs to the vector states before any mixing is taken into account
and $R$ is a parameter combination of order one \cite{Cacciapaglia:2024wdn,Kunkel:2025qld}.
The second type of couplings originates from the mixing of the gluon with $\mathcal V_8$, hence yielding a universal couplings of the massive resonance to quarks:
\begin{align}\label{eq:couplingV8qq}
	\mathcal L_\mathrm{fermions} =-g_s \tan\beta_8\, \bar q\, \slashed {\mathcal V}_8^a t^a_{\mathbf 3} \,q \equiv C_{\mathcal V_8}^{qq} \mathcal O_{\mathcal V_8}^{qq}\,,
\end{align}
The third type is generated by the coupling of the spin-1 resonances to the baryons 
\cite{Erkol:2006sa,Aliev:2009ei} that mix to top and bottom quarks via the partial compositeness mechanism. 
Although the couplings generated by the strong dynamics are inherently vector-like, the chiral mixing 
of the physical states generates chiral couplings to the mass eigenstates leading to 
\begin{align}\label{eq:LagPC}
\mathcal{L}_\text{PC} = 
    \bar{t} \slashed {\mathcal V}_8^a t^a_{\mathbf 3} \left( g_{\rho t t, LL} P_L + g_{\rho t t, RR} P_R \right) t 
    +  \bar{b} \slashed {\mathcal V}_8^a t^a_{\mathbf 3} \left( g_{\rho b b, LL} P_L  \right) b \,.
\end{align}
Note that these couplings are of order $\tilde{g}$, while the chiralities are distinguished by the 
different mixing angles from partial compositeness. Last but not least, note that, while the first two 
types of couplings are completely determined by the chiral Lagrangian, the third one is more model 
dependent. The value of the couplings depend on the quantum numbers of baryons that mix with the 
elementary top and bottom fields, and on the value of the mixing angles. Hence, they cannot be 
predicted in a model-independent way and we will leave them as free parameters. The relevant parameters
of the chiral Lagrangian can be chosen as
\begin{align}
	\tilde g, \quad g_{\rho\pi\pi}, \quad M_{\mathcal V_8}, \quad \xi = \frac{M_{\mathcal A}}{M_{\mathcal V_8}}
	\quad \text{ and } \quad f_\chi\,,
\end{align}
with $f_\chi$ being the decay constant of the color charged pNGBs. $M_{\mathcal A}$ is the mass of
the axial-vectors.

\section{Results}

We will assume in the following a lattice and QCD inspired mass hierarchy, where the color charged 
top-partners are heavier than the spin-1 states, which are heavier than the pNGBs. As mentioned above,
the quantum numbers of the lightest coloured resonances are given in terms of the underlying
hyperfermions for the twelve models considered. They are listed in table~\ref{tab:spectrumQCD} and 
can be group into five classes. 
We focus in the following on the color octet $\mathcal V_8$, which can be singly produced at the LHC, and 
present limits due to existing LHC data.
For this purpose we have first to recall the properties of the coloured pNGBs, which appear as intermediate
decay products. Their phenomenology has been studied in detail in 
refs.~\cite{Cacciapaglia:2015eqa,Belyaev:2016ftv,Cacciapaglia:2020vyf} and we summarize  here only  
their main features.

The ubiquitous colour octet pNGB has two types of couplings: a coupling to gauge bosons generated by 
a topological anomaly and one to top quarks\footnote{The corresponding one to bottom quarks is suppressed by 
the ratio $m_b/m_t$ and, thus, is neglected here.} generated by partial compositeness 
\cite{Belyaev:2016ftv,Cacciapaglia:2020vyf}. Their origin is rather different in nature and, thus,
their relative importance is rather model dependent. The decays of the other colored charged pNGBs 
depend crucially on the scenario. Following the classification in Table~\ref{tab:spectrumQCD}, 
we distinguish four cases:
\begin{alignat}{2}
	&\text{C1}: && \qquad \pi_8 \to t\bar t,\, gg ; \,\, \pi_6 \to bb \,,\\
	&\text{C2}: && \qquad \pi_8 \to t\bar t,\, gg ; \,\, \pi_6 \to tt\,, \\
	&\text{C3}: && \qquad \pi_8 \to t\bar t,\, gg ; \,\, \pi_3 \to \bar b\bar s \text{ or } t\bar \nu,\, b\tau^+\,,\\
	&\text{C4-5}: && \qquad \pi_8 \to t\bar t,\, gg\,.
\end{alignat}
In scenario C2, the sextet has baryon number $2/3$ and electric charge $4/3$. Therefore, partial compositeness 
leads to an unsuppressed coupling to two right-handed top quarks \cite{Cacciapaglia:2015eqa}. 
In case of scenarios C1 and C3, the sextet and triplet have baryon number $\mp 1/3$ and charges $\mp 2/3$, 
respectively. Consequently, they are in general not allowed to decay into standard model fermions by partial 
compositeness alone\footnote{However, for model M5 there are conceivable scenarios 
\cite{Cacciapaglia:2021uqh} in which the lightest color neutral spin-1/2 resonance $\tilde b$ could be 
lighter than $\pi_3$. This would lead to signatures like in supersymmetric models as $\tilde b$ is stable.
Here $\pi_3$ has similar properties as a scalar top quark which decays either in $\tilde b t$ or 
$\tilde b W^+ \bar{b}$ 
\cite{Porod:1996at,Porod:1998yp,Boehm:1999tr}. However, such scenarios requires an extra study beyond the 
scope of this contribution.}. 
Their decays must, therefore, be generated by specific operators that need to violate either baryon number 
or lepton number leading to the decay channels listed above. The di-quark final states $bb$ and 
$\bar{b}\bar{s}$ violate baryon number by one unit, $\Delta B=1$. For $\pi_3$ decays to a quark and 
a lepton\footnote{Effectively this is a third generation leptoquark like the ones discussed in \cite{Faber:2018qon,Faber:2018afz} or R-partiy violating decays of a scalar top quark in supersymmetric models \cite{Restrepo:2001me}.} can be envisioned, violating lepton number by one unit, $\Delta L=1$, which can be generated 
in some models extending partial compositeness to leptons \cite{Cacciapaglia:2021uqh}. Note, that
while the $B$ or $L$ violating couplings can be rather small,  they provide the only decay channel for 
the corresponding pNGB. We assume here that these couplings are sufficiently large to allow for prompt decays.

The colour octet $\pi_8$ can be searched at the LHC via QCD pair production. We assume here that the top 
couplings dominate leading to a four tops final state. A recent reinterpretation \cite{Darme:2021gtt} 
of a CMS search \cite{CMS:2019rvj} leads to a conservative lower bound of $1.25$~TeV for its mass. We note
for completeness, that in C2 models, the contribution of $\pi_6$ can push this limit further up. 
Dedicated searches also exist for the leptoquark decays of the triplet in C3 models, where both 
$b\tau$ and $t\nu$ final states have been searched for by ATLAS and CMS 
\cite{ATLAS:2021jyv,CMS:2020wzx,CMS:2023qdw,ATLAS:2023uox,ATLAS:2024huc} yielding bounds in the range
$1.25-1.46$~TeV, depending on the relative size of the branching ratios. The bounds on this mass
are significantly lower of about $770$~GeV if $\pi_3$ decays dominantly into light quarks 
\cite{ATLAS:2017jnp,CMS:2022usq}.

\begin{figure}[t]
    \centering
    \includegraphics{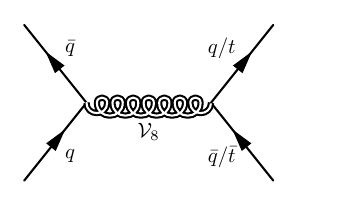}
    \includegraphics{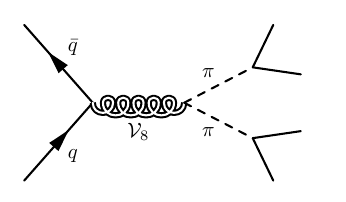}
    \caption{Feynman diagrams of $\mathcal V_8$ single production and decay into quarks or pNGBs.}
    \label{fig:feynmanv8single}
\end{figure}

In the following we consider single production of $\mathcal V_8$ with subsequent decays either into
two quarks or into two pNGBs as shown in fig.~\ref{fig:feynmanv8single}.
The possible decay modes of $\mathcal V_8$ can be classified as follows:
\begin{alignat}{2}
	&\text{C1-2}: \qquad &&\mathcal V_8 \to q\bar q,\; b\bar b,\; t\bar t,\; \pi_8 \pi_8,\; \pi_{6} \pi_{6}^c,\\
	&\text{C3}: &&\mathcal V_8 \to q\bar q,\; b\bar b,\; t\bar t,\; \pi_8 \pi_8,\; \pi_{3} \pi_{3}^c, \\
	&\text{C4-5}: &&\mathcal V_8 \to q\bar q,\; b\bar b,\; t\bar t,\; \pi_8 \pi_8 \,.
\end{alignat}
Scenarios C1 and C2 are distinguished by the decays of $\pi_6$.
The decays into light quarks $q=u,d,c,s$ feature flavour-independent branching ratios, 
while bottom and top quark channels receive additional contributions from partial compositeness, 
see eq.~\eqref{eq:LagPC}, leading to different branching ratios. 
Finally, the relative strength of the pNGB channels is determined purely by colour factors
\begin{equation}\label{eq:pngb_brs}
    \frac{\text{BR} (\pi_6 \pi_6^c)}{\text{BR} (\pi_8 \pi_8)} = \frac{10}{3}\,, \qquad \frac{\text{BR} (\pi_3 \pi_3^c)}{\text{BR} (\pi_8 \pi_8)} = \frac{2}{3}\,.
\end{equation}
Here we have assumed that their masses are equal.

\begin{figure}
    \begin{center}
     SU(6)/SO(6) \\[1mm]
    \includegraphics[trim={0 0 5.1cm 0}, clip, height=0.31\linewidth]{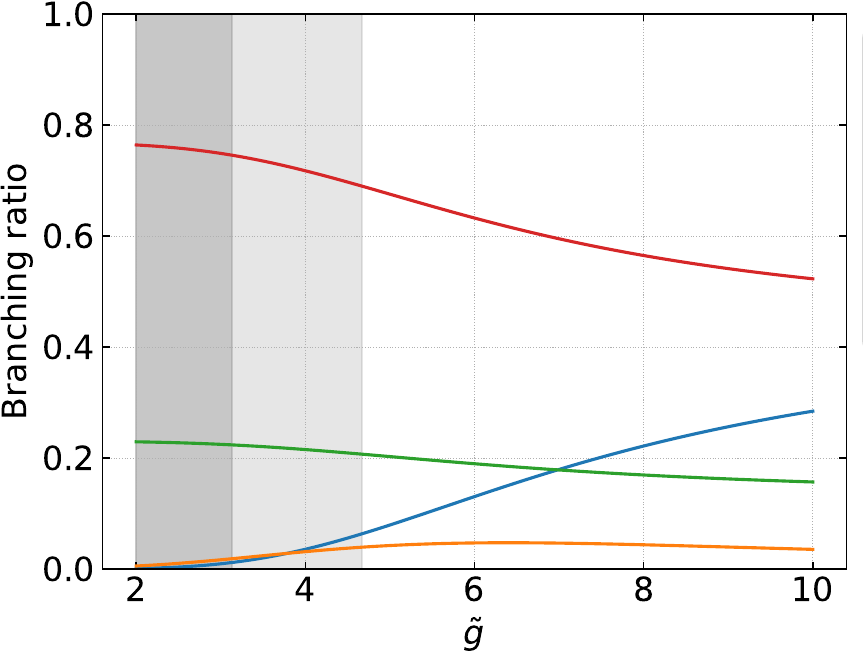} \hfill
    \includegraphics[height=0.31\linewidth]{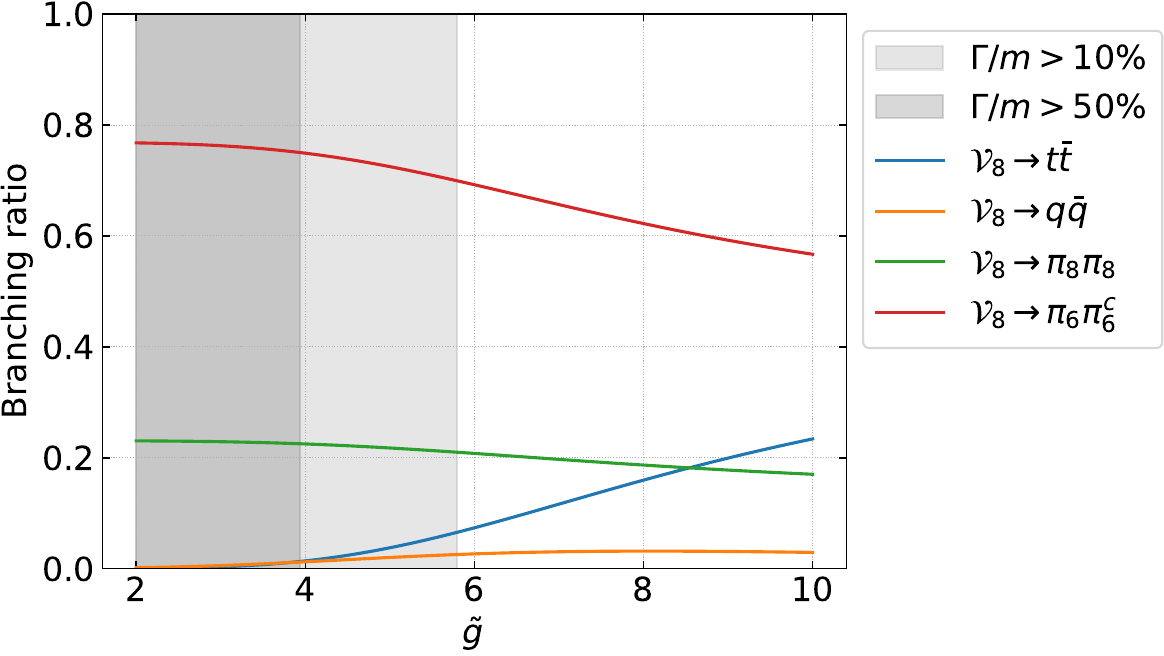} \\[2mm]
     SU(6)/Sp(6) \\[1mm]
    \includegraphics[trim={0 0 5.1cm 0}, clip, height=0.31\linewidth]{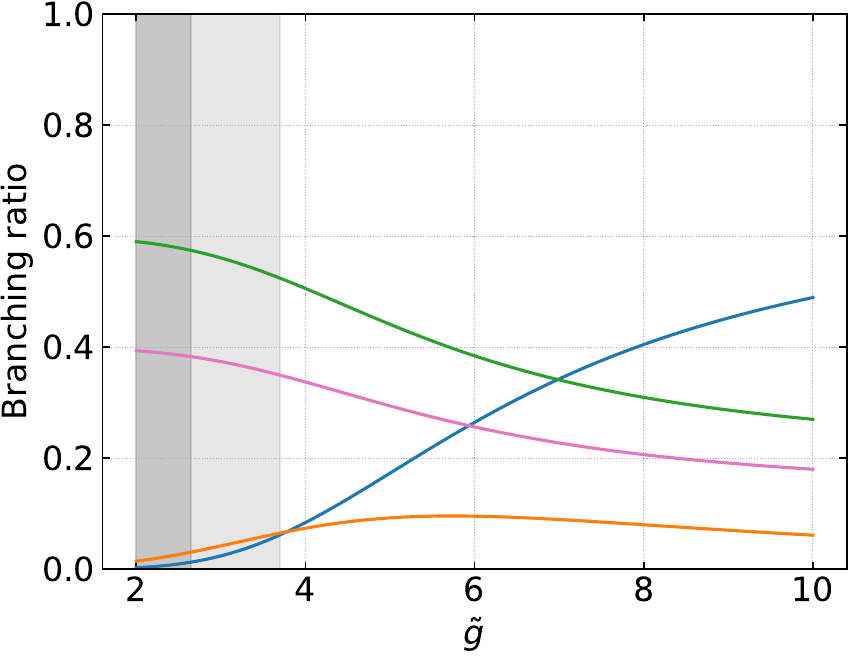} \hfill
    \includegraphics[height=0.31\linewidth]{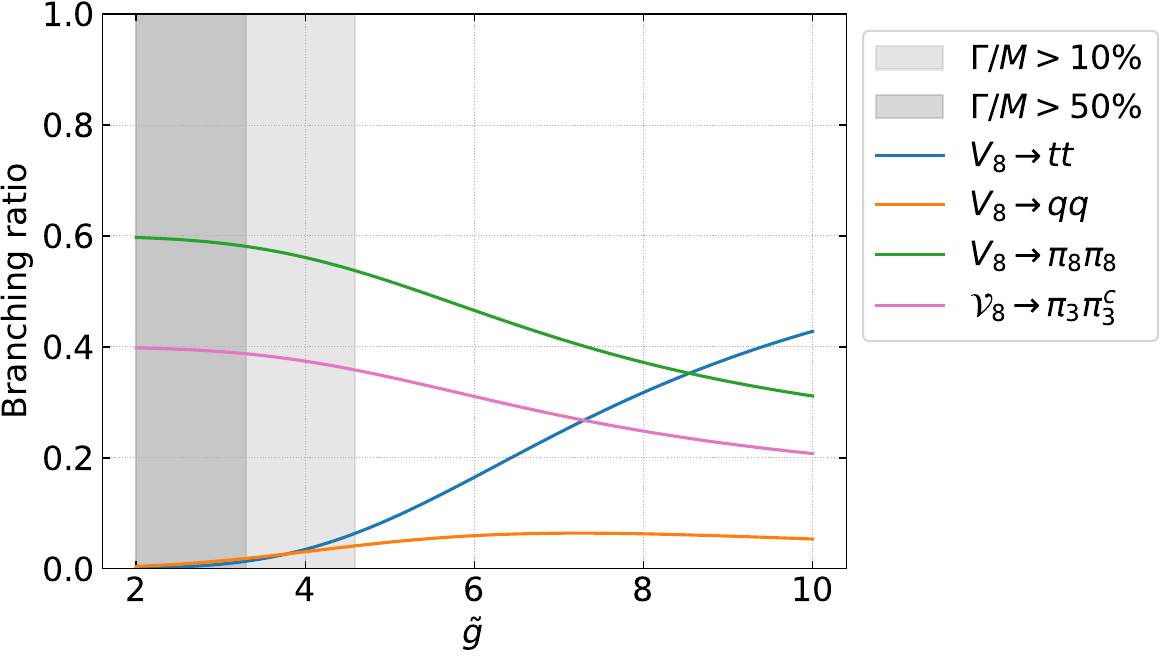} \\[2mm]
     SU(3)$\times$SU(3)/SU(3) \\[1mm]
    \includegraphics[trim={0 0 5.1cm 0}, clip, height=0.31\linewidth]{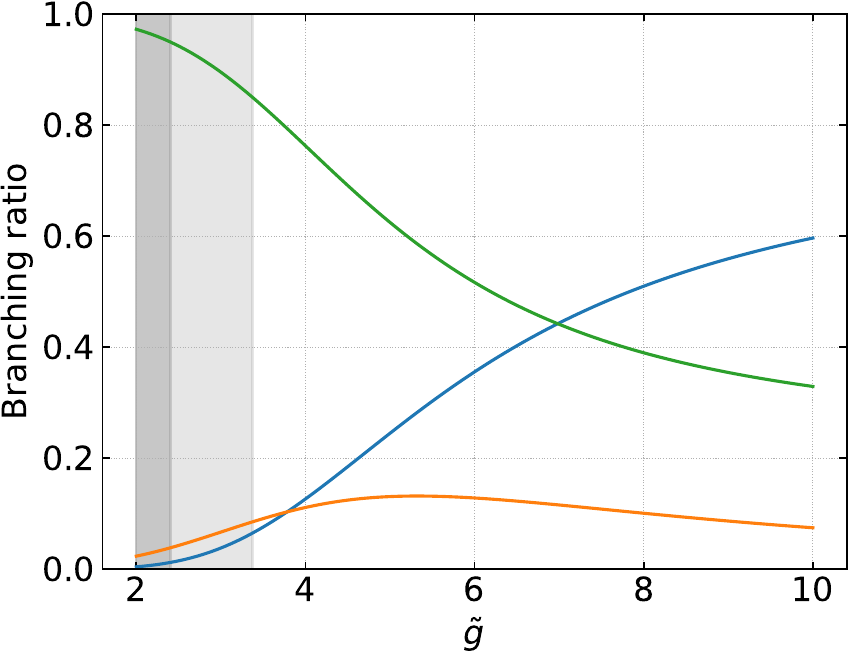} \hfill
    \includegraphics[height=0.31\linewidth]{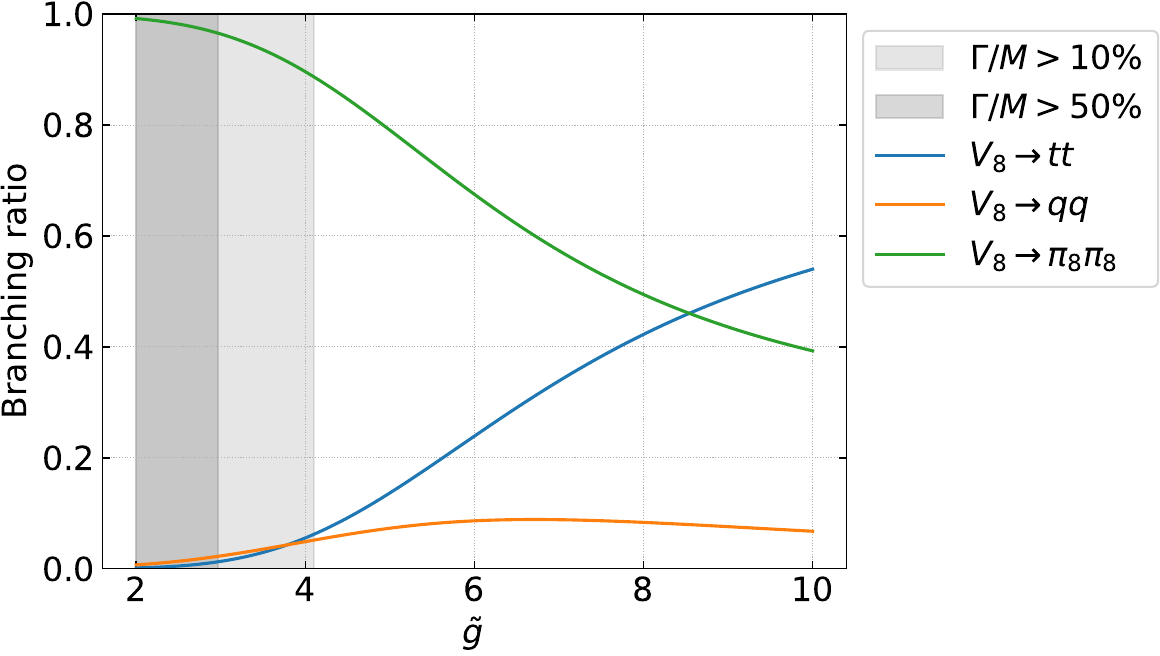} 
    \end{center}
    \caption{${\mathcal V_8}$ branching ratios for the various cosets as a function of $\tilde g$ for $M_{\mathcal V_8}=4.5$~TeV, $g_{\rho \pi \pi}=1$, $m_\pi = 1.4$~TeV, $f_\chi=1$~TeV and fixing the coupling to top quarks to 1. We have fixed in the left column $\xi=1$ and in the right column $\xi=1.4$.
     Moreover, $q=u,d,s,c,b$.}
    \label{fig:branchingratios}
\end{figure}

The importance of each channel depends on the parameter space as can be seen in 
fig.~\ref{fig:branchingratios} which provides some benchmarks for the three cosets as illustration.
The various features can be understood as follows: the partial width into light quarks is controlled by 
the mixing angle $\beta_8$ to gluons which decreases for increasing $\tilde{g}$. The partial width to 
pNGBs receives a dominant contribution proportional to $g_{\rho\pi\pi}$. Here the dependence on 
$\tilde{g}$ is such that this partial width also decreases for increasing $\tilde{g}$. 
For very small $\tilde{g}$, instead, the second term in eq.~\eqref{eq:cv8} starts becoming relevant. This
explains the drop in the $qq$ branching ratio observed in fig.~\ref{fig:branchingratios}. 
Moreover, the scaling in $\tilde{g}$ also explains why the total width of $\mathcal{V}_8$ increases 
for small values of $\tilde g$. We have indicated the regions where the total width becomes large by 
different grey shadings. Finally, the branching ratio to top (and bottom) receives a dominant 
contribution from partial compositeness by our choice of parameters. Thus, they dominate for large values
as they do not scale with $\tilde{g}$.

\begin{figure}
	\begin{center}
	    \includegraphics[width=0.7\linewidth]{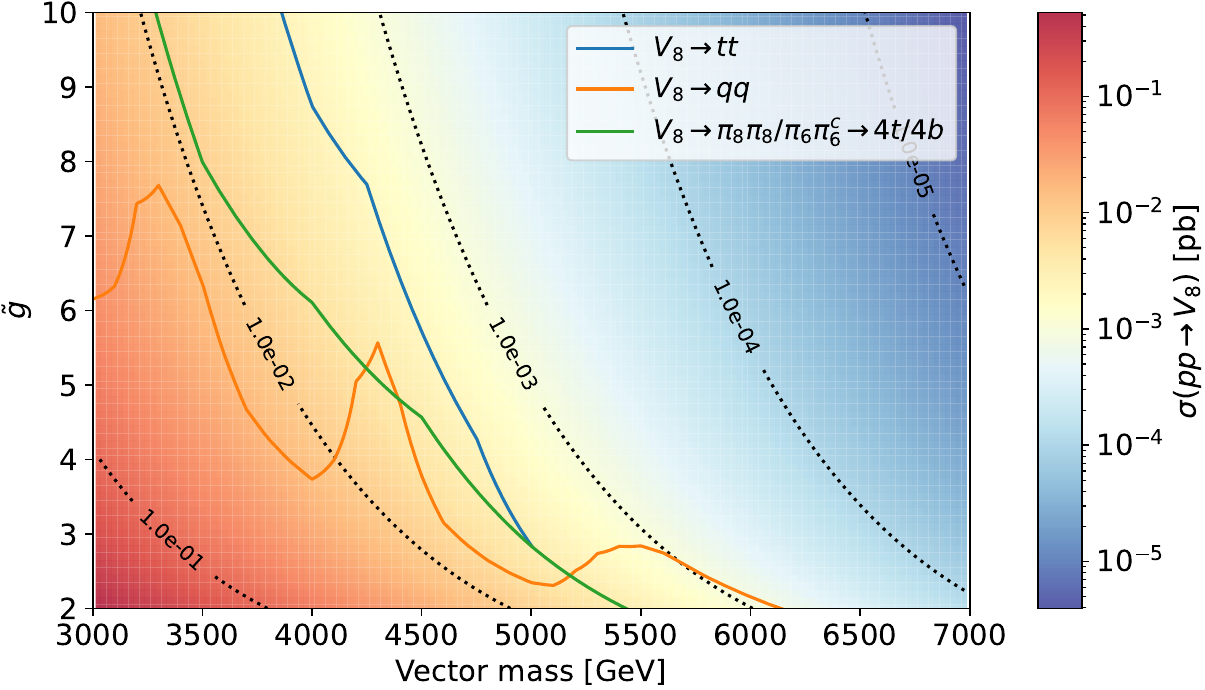} \\[3mm]
	    \includegraphics[width=0.7\linewidth]{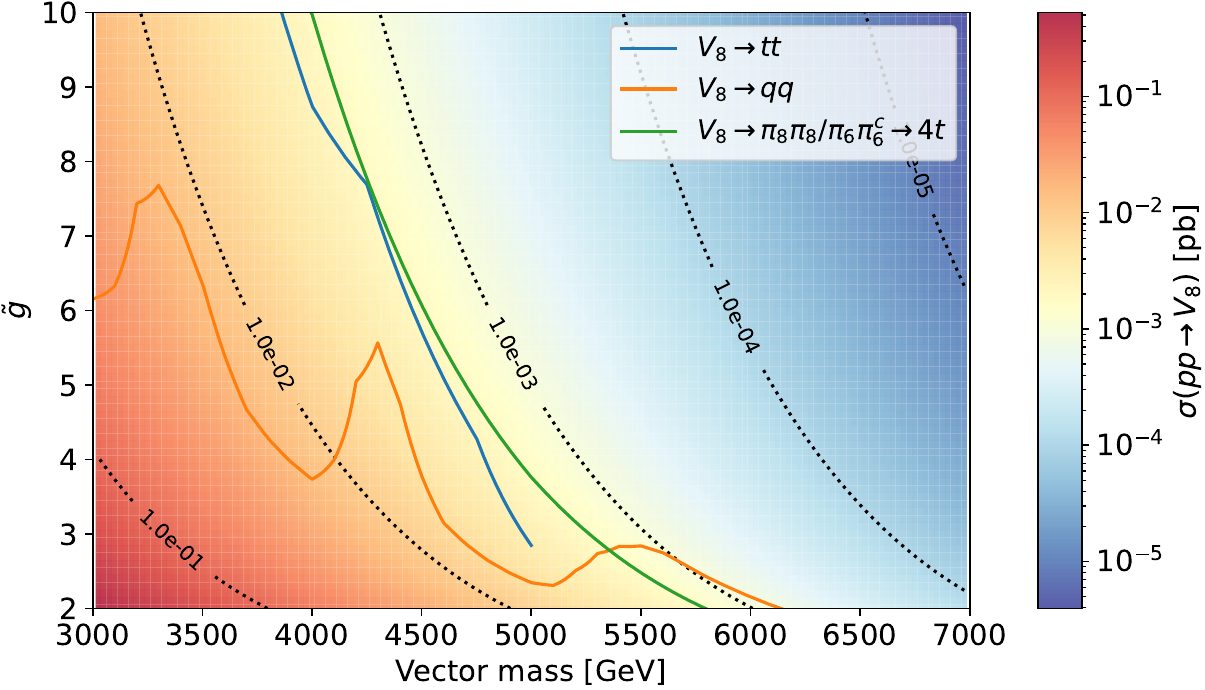}    \\[3mm]
	   \includegraphics[width=0.7\linewidth]{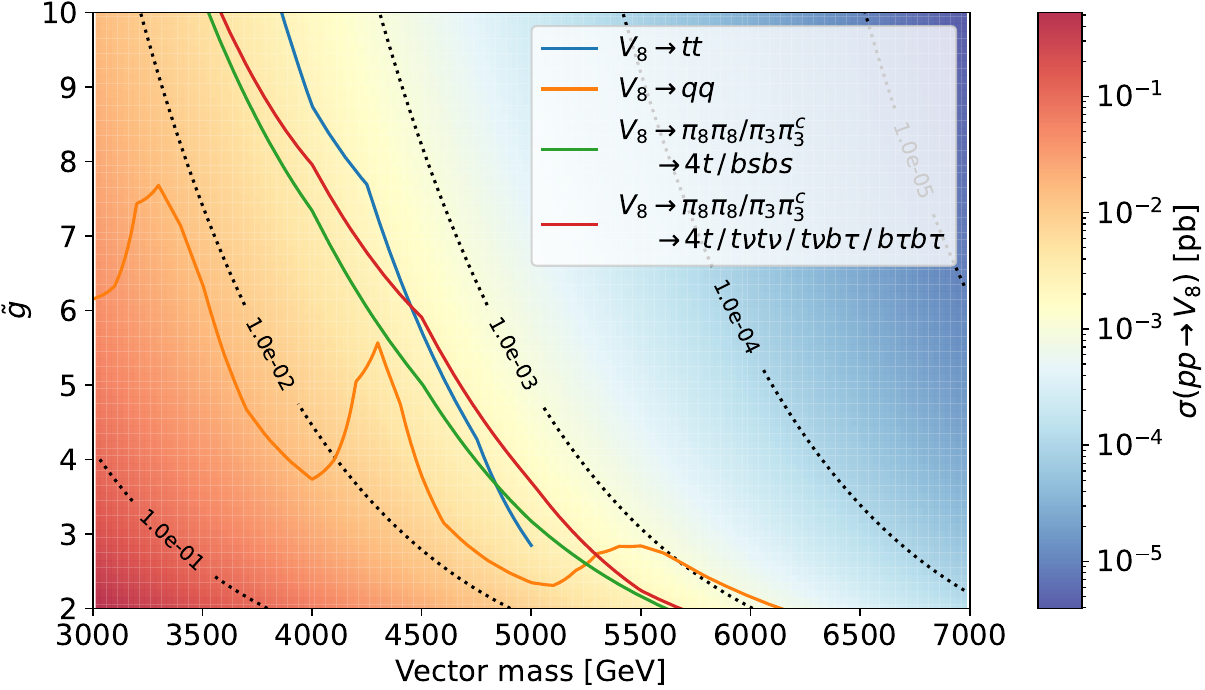}     
	\end{center}    
	\caption{\label{fig:massbounds} Bounds on vector octet single production for the model classes 
	C1 (top), C2 (middle) and C3 (bottom) defined in table~\ref{tab:spectrumQCD}. 
	The colors and the dotted contours indicate the single production cross section. The region to the
    left and below the coloured lines is excluded. The bounds are determined assuming 100\% branching 
    ratio into the indicated channel. For the decays into pNGBs, the ratios of branching ratios in
     eq.~\eqref{eq:pngb_brs} are taken into account. }
\end{figure}

The couplings of ${\mathcal V_8}$ to the light quarks allows for single production, 
see fig.~\ref{fig:feynmanv8single}. The current bounds on the colour octet mass crucially depend on the
branching ratios in the above discussed three channels: light quarks, pNGBs and top quarks. They are
controlled by different couplings. As a model-independent estimate of the bounds we, therefore, 
decided to show limits assuming 100\% branching ratios in the three channels, as shown in
fig.~\ref{fig:massbounds} for the scenarios C1, C2 and C3. The bounds in scenario C4-5 are very similar
to the ones of scenario C2, see refs.~\cite{Cacciapaglia:2024wdn,Kunkel:2025qld} to which we refer for 
further details. The bounds are extracted from the following searches:
\begin{itemize}
    \item di-jet: search for high mass di-quark resonances \cite{CMS:2019gwf};
    \item di-top: search for $t\bar t$ resonances \cite{ATLAS:2020lks};
    \item pNGBs: recasts of SUSY searches \cite{ATLAS:2019fag,ATLAS:2021twp,ATLAS:2022ihe} 
               implemented in \texttt{CheckMATE}  \cite{Drees:2013wra,Dercks:2016npn}.
\end{itemize}
The colours indicate the Drell-Yan type cross section, which depends only on $\tilde{g}$ and 
$M_{\mathcal V_8}$. The region below and to the left of the lines is excluded. The results show that 
the limits on the mass are roughly the same in the range of $4$ to $5$~TeV for all scenarios. 
Note however, that the region for small $\tilde{g} \lesssim 3$ cannot be fully trusted as it corresponds to
widths above 50\% of the mass and, thus, a dedicated analysis would be required which however is beyond
the scope of this contribution. 

\section{Discussion and outlook}

We have presented the LHC phenomenology of spin-1 resonances carrying QCD charges in Composite Higgs Models
with a focus on models which allow for fermionic UV completions \cite{Ferretti:2016upr,Belyaev:2016ftv}.
More precisely, we have worked out their properties for three types of cosets, SU(6)/SO(6), 
SU(6)/Sp(6) and SU(3)$\times$SU({3)/SU(3) including their 
most relevant production at the LHC and their decay channels. These are symmetric cosets and, thus,
they contain two sets of spin-1 resonances: vector states that coupling to two or more pNGBs and 
axial-vector states which have to couple to at least three pNGBs.

All scenarios feature a vector $\mathcal{V}_8$ which mixes with the gluons. As a consequence, 
$\mathcal{V}_8$ can be singly produced at hadron colliders via Drell-Yan, whereas all other states can 
only be pair-produced. The $\mathcal{V}_8$ can either decay into a quark pair or into two pNGBs leading 
in all cases to the final states $q\bar{q}$ ($q\ne t$), $t\bar{t}$ and $4t$. We have investigated
the bounds on $\mathcal{V}_8$ from current LHC data. For this we have used on the one hand direct searches
for resonances decaying to two SM quarks. In the other hand, we have used recasts from searches for
supersymmetric particles for cascade decays via pNGBs. We have found in all scenarios bounds on 
its mass ranging from  $3.5$~TeV to $6$~TeV. 
However, we note for completeness that in some part of parameter space these results has to
be taken with a grain of salt as this resonance has a rather broad total decay width which calls for 
a detailed reanalysis of the data.

The high luminosity run at the LHC will certainly allow to further improve the mass limits on 
$\mathcal{V}_8$. However, already the current bounds in Fig.~\ref{fig:massbounds} imply that
pair production of all spin-1 resonances will have rather small cross sections at the LHC, hence 
making their detection unlikely. Thus, we briefly sketch here their signatures from pair 
production at a 
future high energy hadron collider, e.g.~a prospective 100 TeV pp collider, which will be presented
more detail in
a future contribution. However, we should stress that single production of $\mathcal{V}_8$ remains the 
leading discovery channel and, thus, one would expect to discover $\mathcal{V}_8$ before pair 
production becomes relevant.

The sextets, which feature the largest pair production cross sections \cite{Cacciapaglia:2024wdn}, are present in scenarios C1, C2 and C3. Their decays can be classified as follows:
\begin{eqnarray}
  \mathcal{A}_6 &\to& \pi_8 \pi_8 \pi_6\ (t\bar{t}t\bar{t}bb)\; \text{ and } \; \pi_6 \pi_6^c \pi_6 \ (\bar{b}\bar{b}bbbb)\quad \text{in C1}\,,\\
        \mathcal{A}_6 &\to& tt\; \text{or} \; \pi_8 \pi_8 \pi_6\ (t\bar{t}t\bar{t}tt)\; \text{ and } \; \pi_6 \pi_6^c \pi_6 \ (\bar{t}\bar{t}tttt)\quad \text{in C2}\,,\\
        \mathcal{V}_6 &\to& \pi_8 \pi_3^c \to (t\bar{t}) (\bar{b}\bar{s}\; \text{or}\; q l) \quad \text{in C3}\,.
\end{eqnarray}
The two-body decay of $\mathcal{V}_6$ in C3 is driven by $g_{\rho\pi\pi}$, hence is will likely imply 
a large decay width. By contrast, thanks to the three body final state of $\mathcal{A}_6$ in pNGBs, one
expects the widths of $\mathcal{A}_6$ to remain small compared to its mass. 
In scenarios C2, $\mathcal{A}_6$ can also decay into two tops which however is 
suppressed by $v^2/f^2 \leq 0.04$ (for $f \geq 1$~TeV). At the same time the three-pNGB channel is suppressed 
by a phase space factor and, thus, one expects that both decay modes will compete and that 
the widths remain small for these states. In both cases, C1 and C2, the pair production of $\mathcal{A}_6$
will lead to final states containing several top and bottom jets.

The axial colour octet $\mathcal{A}_8$ also has a sizeable pair cross section but somewhat smaller
than the ones of the sextets. Its leading decay channel is into two tops in all cases. Thus, one
has a four-top final state with potentially large width effects.

Scenarios C1, C2 and C3 features in addition the colour triplets which have cross sections which
are roughly an order of magnitude smaller if all spin-1 resonances have about the same mass.
They have  the following decays:
\begin{eqnarray}
 \mathcal{V}_3 &\to& \pi_8 \pi_6^c\ (t\bar{t}\bar{b}\bar{b})\; \quad \text{ in C1}\,,\\
 \mathcal{V}_3 &\to& \bar{t}\bar{t}\; \text{or} \; \pi_8 \pi_6^c\ (t\bar{t}\bar{t}\bar{t})\quad \text{ in C2}\,,\\
 \mathcal{A}_3 &\to& \pi_8 \pi_8 \pi_3\; (t\bar{t}t\bar{t}+\bar{b}\bar{s}\; \text{ or }\; q l)\; 
 \text{ or }\; \pi_3 \pi_3^c \pi_3 \; (\bar{b}\bar{s}bs\bar{b}\bar{s}\; \text{or}\; ql\bar{q}lql)\quad \text{ in C3}\,.
\end{eqnarray}
$\mathcal{V}_3$ will mainly decay into $\pi_8\pi_6^c$ independent of the scenario as the di-top coupling 
in C2 is suppressed by $v/f$. One expects a large total decay width in most of the allowed parameter space.
In contrast, the $\mathcal{A}_3$ in C3 will be a narrow resonance with interesting final states rich 
in top quarks and possibly leptons, if the $\pi_3$ decays violate lepton number.

\acknowledgments
I thank  G.~Cacciapaglia, A.S.~Cornell, A.~Deandrea, T.~Flacke, M.~Kunkel and R.~Str\"ohmer for interesting and fruitful discussions. 
I thank M.~Kunkel for creating the plots in fig.~\ref{fig:branchingratios}.
I also would like to thank the organizers of the Corfu workshops for creating such a stimulating environment.
This work has been supported by DFG project  nr PO/1337-12/1.

\end{document}